\documentclass[twocolumn,aps,prl,english,twocolumn,epsfig,groupedaddress]{revtex4}
\usepackage{graphicx}
\usepackage[latin1]{inputenc}
\begin{document}
\title{\bf Non-perturbative  $J_{pd}$ model and ferromagnetism  in dilute
magnets }

\author{Richard ~Bouzerar$^{1}$\footnote[4]{email: richard.bouzerar@u-picardie.fr} Georges ~Bouzerar$^{2}$\footnote[6]{email: georges.bouzerar@grenoble.cnrs.fr} and Timothy ~Ziman$^{3}$\footnote[5]{and CNRS, email: ziman@ill.fr} 
}

\affiliation{$^{1}$Universit\'e  de Picardie Jules Verne, 33 rue Saint-Leu, 80039 Amiens Cedex 01\\ 
$^{2}$Laboratoire Louis N\'eel 25 avenue des Martyrs, CNRS, B.P. 166 38042 Grenoble Cedex 09
France.\\   
$^{3}$Institut Laue Langevin B.P. 156 38042 Grenoble 
France.\\
}            
\date{\today}

\begin{abstract}
We calculate magnetic couplings in  the   $J_{pd}$ model for dilute
magnets,  in order both to  identify the relevant parameters which 
control ferromagnetism and also to bridge the gap between first principle calculations
and  model approaches. 
The magnetic exchange interactions are calculated non-perturbatively and 
disorder in the configuration of impurities is treated exacly, allowing us to test  the validity
of effective medium theories.
 Results  differ qualitatively    from those of  weak coupling. In contrast to mean field theory, increasing $J_{pd}$ may not favor high Curie
temperatures: $T_C$    scales
primarily with the bandwidth. 
High temperature ferromagnetism at small dilutions is  associated with resonant
structure in the p-band. Comparison to diluted magnetic semiconductors
indicate that  Ga(Mn)As has such a resonant structure and thus   
this material is already close to optimality.  
\end{abstract}
\parbox{14cm}{\rm}
\medskip

\maketitle

\section{}
The basis of our understanding of magnetism
in  
 metals is the
 Ruderman-Kittel-Kasuya-Yosida (RKKY) interaction\cite{RKKY}.
The magnetic exchange results from  interaction of  inner d- or f- 
localized orbitals  and the outer s- or p- that form the conduction bands.
In the standard form it is calculated   with perturbative treatment
of the coupling between the moments and the carriers on  
a definite Fermi surface, which is  itself weakly affected by the presence of the magnetic moments.
This leads to successful interpretation of oscillating exchanges in, for example,
transition or rare earth metals.
However the same concepts are often applied in situations where
 this perturbative approach  is not really applicable.
There are many situations where the coupling
between moments and itinerant carriers is not weak and it is important
to determine changes in magnetic exchange in this regime,
for example in the field of manganites\cite{MillisManganite}, double perovskites\cite{Perovskite} or in materials
modelled by Kondo lattices\cite{Kondo}.
In another  field of great current interest, diluted magnetic semiconductors (DMS), 
a belief, incorrect in our view\cite{Bouzerar2}, in the applicability of weak coupling has resulted
from  an unhappy juxtaposition
of two approximations: a  simplified RKKY view of magnetic interactions
and over-simplified molecular field theory for the thermodynamics.
This  leads to the estimate for the Curie temperature $T_C^{MF}=J_{pd}^{2}xn^{1/3}/t$, 
where $t=1/m^{*}$  is the inverse of the effective mass. $J_{pd}$ is the  
coupling between d-orbital moments and the primarily p-orbital
conduction carriers, $x$ the concentration of d-orbitals and $n$ the 
density of carriers.
This estimate has even 
been influential in the search for new materials:
attempts to  increase $T_c$  at low doping have focussed  on seeking large $J_{pd}$ and small
bandwidth (large effective mass).
We shall show that the widely used
$T_C^{MF}$ is in fact very misleading in the non-perturbative limit.
Apart
from potential applications \cite{Ohno}, these materials  are of fundamental interest for exploring
ferromagnetism in a regime where geometric disorder is strong (because of the low concentrations of dopants) and  the
values of $J_{pd}$ are strong. 
Unfortunately these   issues of principle have
been obscured by insistence \cite{Prospects} 
on the weak-coupling picture,
and has hidden  richer underlying aspects. 
We  argue that these systems  are particularly interesting
because the ferromagnetism, 
can {\it only} be understood if the crucial coupling $J_{pd}$ 
is treated non-perturbatively. 
An approach  taking  {\it ab initio} 
estimates of couplings and reliable treatment of the thermodynamics
\cite{Bouzerar2} can be 
successful,  but the  effective magnetic interactions found, which include
a multitude of effects, have not 
been understood in simple terms.  Calculations  must be taken
for each individual composition.
A more systematic approach
for finding promising new materials is possible if we have    
simple but accurate model
systems.
We can then identify which parameters 
are important:  $J_{pd}$, the carrier density
or the bandwidth of the conducting band? We can also examine
assumptions
of the {\it ab initio} approach: e.g. the Coherent Potential Approximation (CPA) for   
carrier disorder. 
The method we shall employ makes 
effective medium approximations {\it  neither} in the deriving of the 
magnetic couplings {\it nor} in calculating magnetic properties.
Thus an aim is to make a  bridge
between {\it ab initio}  and  conceptually simpler
model approach. 
A practical question is whether $T_C$ can be increased
by choice of other chemical dopants. 
In lightly doped III-V semiconductors,  it will appear that
Ga(Mn)As is in fact  close to  optimal.
Here  we discuss  the theory  appropriate to dilute magnets 
where  the dopants provide well defined moments,
but the concepts are relevant to  a wider class of 
novel magnets ( ``$d^0$'' magnets)  in which there is no clear distinction between localized magnetic moments
and the itinerant carriers of the (doped) host\cite{CoeyNature,Dzero}. 

\par
The $J_{pd}$ model (or dilute Kondo lattice) is of 
 a band of atomic d- orbitals coupled to an itinerant ``p-band'' of  carriers via
a  Hund's rule coupling. In the manganites, for example,  the ``p'' band is in fact
be constructed from $e_g$ orbitals. 
\begin{equation}
H=-\sum_{ij, \sigma} t_{ij}c_{i\sigma}^{\dagger}c_{j\sigma}+
\sum_{i}J_{i} {\bf S}_{i}\cdot {\bf s}_{i}+
\sum_{i \sigma}V_{i} c_{i\sigma}^{\dagger}c_{i\sigma}
\end{equation}
$t_{ij}=t$ for $i$ and $j$ nearest neighbors and zero otherwise.
In the exchange between localized impurities spin and itinerant
electron gas $J_{i}$ is a random variable: $J_i=J_{pd}$ and $V_i=V$
 if the site $i$ is occupied by a magnetic impurity, and zero otherwise. 
$V_i$ is here to mimic the effect of chemical substitution that accompanies
the presence of the magnetic moment. ${\bf S}_{i}$ is the magnetic impurity spin operator at site $i$ and ${\bf
s}_{i}=c_{i\alpha}^{\dagger} (1/2\mbox{\boldmath$\sigma$}_{\alpha
\beta})c_{i\beta}$ the spin operator,  at the same site, of the
itinerant electron gas. The d spins are distributed randomly at low concentration $x$ on a regular
lattice, for simplicity taken to be simple cubic.
We remark that this  model can be extended trivially to 
include several bands, as would be appropriate for models of double perovskites or manganites,
where both the hopping $t_{ij}$ and on-site potentials $V_i$ may be random\cite{MotomeFurukawa, Pinaki}.
The  results that will follow are non-perturbative
in the $J_{pd}$ coupling and exact in the disorder.
A crucial  effect that will be seen, and is neglected to low
order in perturbation theory, is that as  $J_{pd}$ increases, the conduction band is increasingly affected
by the presence of the magnetic ions. 
The calculations are made in the  ferromagnetic
phase, whose stability is tested {\it a posteriori}, and therefore start from the configuration of spins $S$ that 
are fully aligned in some direction $S_i^z=S$ for all occupied i.
We can diagonalize in the Hilbert space (spins plus electrons) defined by 
the fully polarized d-spins 
provided that we ignore the transverse fluctuations of those spins,
of  order $\frac 1 S$\cite{Bouzerar1}.
Thus our theory is exact in the limit of large $S\rightarrow\infty$
For the electronic degrees of freedom this leaves a Hamiltonian quadratic
in fermion operators. The effect of the d-orbitals is to produce an effective
spin-dependent term  of $V_i^{\pm}=V {\pm}\frac{J_{pd}S}{2}$ on each occupied state, i.e. a random site potential
for up (+)and down (-) carriers. For each configuration of disorder this Hamiltonian
will  be diagonalized exactly. 
\par
\begin{figure}[tbp]
\includegraphics[width=8.cm,angle=-90]{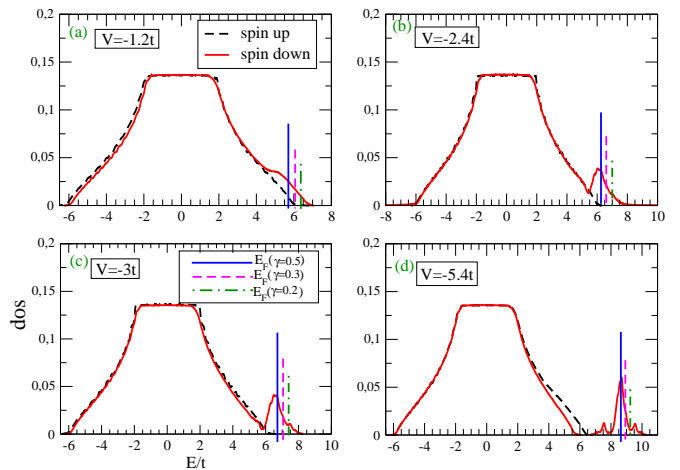}
\caption{(Color online)
Density of states. The  bands of spin down and up are continuous and dashed
respectively. Energies are in units of t, $J_{pd}S=5t$ and the concentration is 5\%. The Fermi energy is indicated for three band fillings $\gamma$}
\label{dos}
\end{figure}

In Fig.\ref{dos} we show the single particle density of states 
for fixed $J_{pd}S=5t$ and concentration $x=5\% $
of magnetic impurities  and different  values of the d-orbital potential $V/t= -1.2, -2.4,-3$ and -5.4.
The parameters were chosen to reproduce the general form seen in doped III-V semiconductors, as will 
be discussed in more detail below. 
For all  calculations presented, the  systems were sufficiently large (typically $\sim 16^3$ sites)
that finite size effects are negligible.
As V increases from the smallest value,
a well defined impurity band for the minority band splits from the valence band.
The Fermi energy is shown for different values of $\gamma=n_h/x$, where $n_h$ denotes the density of holes in a filled band.
We will see below how the magnetic couplings  are correlated  with the form  of the  density of states and the position of the Fermi energy.
\begin{figure}[tbp]
\includegraphics[width=7cm,angle=-90]{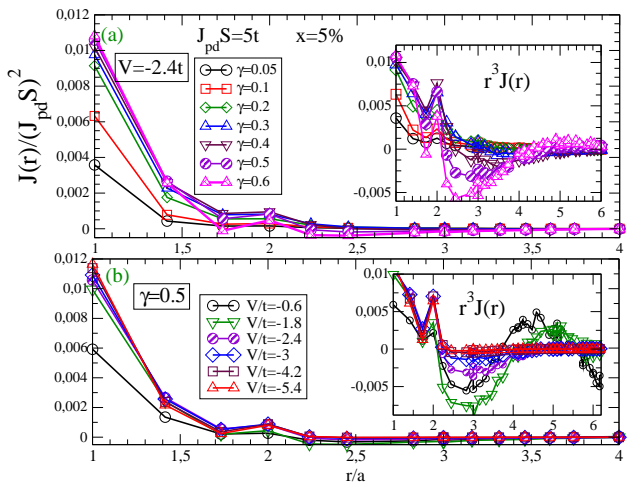}
\caption{(Color online) (a)Magnetic exchange  as a function
of distance, in units of  lattice spacing,
for different values of $\gamma$. Other parameters correspond to case (b) of  Fig.~\ref{dos}.  (b) Exchanges
for different values of the potential  $V$ at half filling. In inset, the exchanges rescaled:  $J(r)r^3$}
\label{couplingspace}
\end{figure}
\par
The magnetic exchange
interactions are derived at all distances between the different moments
from the 
carrier Green functions $G_{{i} {i}^{\prime}}^{\sigma}$\cite{Lichtenstein}.
This  defines an effective (dilute) Heisenberg model for the 
magnetic moments,  again valid for spins  in the classical limit,
${\cal H}_{Heis}=-\sum_{{  i},{  j}}{\cal J}_{{  i},{  j}} S_i\cdot S_j$.
Exchanges are calculated
 by applying infinitesimal fields to the aligned ferromagnetic state\cite{Lichtenstein}.
\begin{eqnarray}
 {\cal J}_{{  i},{  j}}= -{\frac {1} {\pi}} \Im \int_{-\infty}^{E_F} \Sigma_{{  i}}G_{{  i},{  j}}^{\uparrow}(\omega) \Sigma_{{  j}} G_{{  j},{  i}}^{\downarrow}(\omega) d\omega\nonumber
\end{eqnarray}
In this case $\Sigma_i=V_i^{+}-V_i^{-}$.
We note that the  calculations 
use
no effective medium approximation.
In Fig.\ \ref{couplingspace}(a) we show the  couplings  
 averaged
over different impurity configurations. We vary the hole doping per d-orbital $\gamma$ for fixed $x=5$\% and $V=-2.4t, J_{pd}S= 5t$ ( corresponding to  Fig.\ref{dos}(b)). 
These clearly differ from RKKY for all fillings in that they do not
oscillate around a mean value of zero and beyond 3 lattice spacings they essentially vanish. While varying sharply
with distance they do not change in sign, at least for small $\gamma$.
The couplings at short distances increase
monotonically with $\gamma$ but for sufficiently large $\gamma$  
antiferromagnetic values appear at larger distances (see inset).
In Fig.\ \ref{couplingspace}(b) the  $J(r)$ with different potentials $V$ are
shown for fixed $\gamma$. As $|V|$ increases the interactions increase at short distances
but become of very short range. For $|V|$ small the longer range oscillations
appear clearly, as seen in the inset.
Thus from Fig.\ref{couplingspace}  there is  a range of $V$ and $\gamma$ 
where the couplings
are all ferromagnetic ($J(r) \ge 0$). This is associated
(see  Fig. \ref{dos} (b)) with  incipient development of a
visible impurity band just at the band edge.
This ferromagnetic ``bias''  results from the resonant form of the p-band as advanced in Ref.\cite{RichardBouzerar}. 
In that paper  we compared to  exchange calculated in a toy model between two 
resonant states\cite{Caroli},  arguing that  particle-particle channels responsible for super-exchange should be suppressed.
In contrast, in the present calculation 
there is no artificial separation between 
the resonant states and the rest of the conduction band nor between different contributions (particle-hole and particle-particle  channels).
The ideas are coherent in that the $J_{pd}$ acting between
the d- spin and the resonant p-state suppress particle-particle
virtual states.
For the largest values of $|V|$ the clear suppression of exchange
for large distances (see inset to Fig. \ref{couplingspace}(b))
is a precursor to  pure magnetic double exchange type\cite{DoubleExchange}.
For  $J_{pd}$ and $V$ strictly  
infinite we expect the coupling to reduce to nearest neighbours:
${\cal J}_{ij}=t_{ij}\langle c^{\dagger}_ic_j\rangle$\cite{MotomeFurukawa}
For  $J_{pd}S$  finite, however,  the exchange interactions
extend further  than nearest-neighbours\cite{Terakura}.
\par
\begin{figure}[tbp]
\includegraphics[width=7cm,angle=-90]{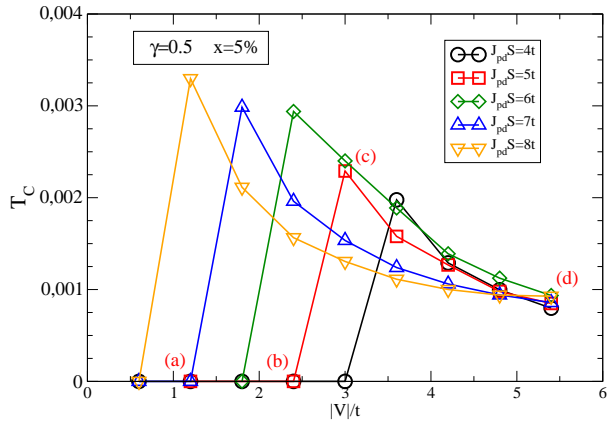}
\caption{(Color online)  Curie temperature in units of t at half-filling as a function of potential V 
and varying coupling $J_{pd}$. Points (a)-(d) correspond to the density of states
of Figures 1(a)-(d) respectively} 
\label{TcV}
\end{figure}

We now calculate the Curie temperatures $T_C$ 
by the  self-consistent local random phase approximation (SC-LRPA)\cite{Bouzerar2}
which treats geometric disorder exactly, as is particularly
important in the dilute limit.
SC-LRPA  has been  successful in reproducing experimental $T_C$ of DMS and agrees with Monte Carlo
simulations\cite{mc1,Sato}. 
\begin{figure}[tbp]
\includegraphics[width=7cm,angle=-90]{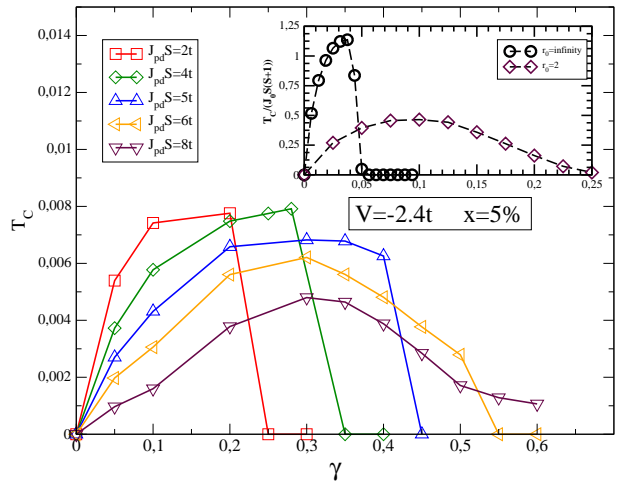}
\caption{(Color online) Curie temperature in units of t  as a function of carrier density. In insert the same calculation for the RKKY couplings see ref \cite{RichardBouzerar} Fig.4}
\label{Tcgamma}
\end{figure}
Figs.~\ref{Tcgamma} and ~\ref{TcV}   shows the Curie temperatures (T$_C$)  
as  functions of  $\gamma$ and  $V$ 
at half-filling respectively, and various values of  $J_{pd}$. Unlike the predictions 
of $T_c^{MF}$, $T_C$ increases monotonically {\it neither} with 
$J_{pd}$ {\it nor} with $\gamma$. 
For $\gamma=0.1$, for instance, $T_C$ {\it decreases} with $J_{pd}$!
Large values
of $J_{pd}$ are primarily useful in  broadening the range
of stability of ferromagnetism with $\gamma$. For comparison
we show in inset previous results
in the RKKY regime\cite{RichardBouzerar}.
At the smallest values of $J_{pd}$ the region of stability (note the 
difference in scale $\gamma$) approaches
the RKKY results, as expected. 
In Fig.~\ref{TcV}  we see that  $T_C$ is less sensitive to  
$J_{pd}$ in the large  $V$ limit. This is essentially
the limit of double exchange. The ferromagnetic bias
is essentially the physics of double exchange\cite{AlexanderAnderson},
and   the scale of $T_C$  is  $t$. 
Note  we  make the assumption 
that  the  ferromagnetic $T_C$  calculated with couplings {\it averaged over
configurations} is equal to that obtained, in the thermodynamic limit,
with {\it unaveraged} couplings. 
We have checked this assumption
and it is accurate.

\par
We can  use our results as a simple model for doped
Ga(Mn)As.
We take parameters from photoemission\cite{PhotoEmission}
$J_{pd}\approx 1.2 eV, S=5/2,J_{pd}S\approx 3eV$.
From the band-structure of GaAs:
$m^{\star}\approx 0.5 m_e$, lattice spacing $a_0=5.65\AA$, this gives $t\approx 0.7eV$ and thus $J_{pd}=4t$.
The value of $V/t=-2.4 eV$ was chosen to mimic the calculated density
of states of Ga(Mn)As.
For these parameters and $x=5\% $, from Figure.~\ref{Tcgamma},  we obtain 
a maximum   $T_C^{1 band}\approx 65K$ at $\gamma= 0.3$. Because there are two degenerate
hole bands and we  have neglected all correlations in the p-bands,
we can simply extend  to 2 independent bands:
$T_C^{2 band}=2 T_C^{1 band} \approx 130K$, close to 
the measured value in annealed samples of Ga$_{.95}$(Mn)$_{0.05}$As.
This is a crude comparison as the lattice is face-centred cubic and not cubic
and, at first sight, the natural concentration to chose for uncompensated samples 
might  be  two half-filled bands $\gamma=0.5$ to give one carrier per dopant. We  argue, however  that  $\gamma=0.3$ {\it is}
comparable to uncompensated Ga(Mn)As since it is at  this  doping
that the  density of states 
(calculated from band structure\cite{Josef}) most resembles  that of Figure\ref{dos}b, with
$E_F$ slightly {\it above} the maximum in the impurity peak. 
The apparently  remarkable agreement for such a simple model
is found for  the range
of concentrations relevant to experiments.
We conclude {\it if}  $J_{pd}$  is treated non-perturbatively
we do 
not need
extensions such as the 6-band Luttinger model\cite{Prospects}
to give accurate quantitative
comparison.
We can also
understood from the position of the impurity band  why  Ga(Mn)As has a larger T$_C$ than Ga(Mn)N and In(Mn)As which
resemble more cases (d) and (a) of Fig. \ref{dos} respectively\cite{Bouzerar2}. 
Since the estimated $J_{pd}$ and bandwidths for these materials
are all very similar, the important difference is the potential
$V$. From the calculated $T_C$s shown in the Figures and 
comparison with the density of states in Fig \ref{dos} (b),we see that Ga(Mn)As has close to  optimal parameters. 
This may explain failure to find higher values of $T_C$ in III-V semiconductors without invoking other effects: local clustering
or increasing bandwidth. 

\par
To conclude,
we have  shown  that in the non-perturbative regime  the 
$J_{pd}$ model has effective magnetic couplings  very different from those of RKKY
and resemble  those of {\it ab initio} calculations for the DMS. 
This similarity  supports
the accuracy  of  CPA as used in the first principle  calculations  for calculating
magnetic exchanges, an approximation  that has often been questioned.
While we have presented results in the dilute regime,
we note that our approach, which  gives the exchange interactions in real space,
can be used in more general situations at higher concentrations
or with inhomogeneities, as may be applicable in  other materials described
by the $J_{pd}$ model and its extensions.
For dilute magnets  a region of parameter space gives  couplings that 
are all ferromagnetic. 
The reason\cite{RichardBouzerar} is made explicit:
optimality  is associated with a  resonant structure seen in the 
p-band. 
The potential $V$ must be  sufficiently strong that this occurs,
but if  it is too strong, couplings are too short-range to
maintain long range order. 
Unlike previous mean field theories ($T_C^{MF}$),
the temperature {\it scale} for ferromagnetism
is controlled {\it not by $J_{pd}$} but by the bandwidth $t$.
The physics is essentially that of double exchange in this regime.
The values of Fermi energy, $J_{pd}$
and the potential shift $V$ 
are important in obtaining optimal couplings.

\end{document}